# Demographic Homeostasis and the Evolution of Senescence

**Abstract.** Existing theories for the evolution of aging and death treat senescence as a side-effect of strong selection for fertility. These theories are well-developed mathematically, but fit poorly with emerging experimental data. The data suggest that aging is an adaptation, selected for its own sake. But aging contributes only negatively to fitness of the individual. What kind of population model would permit aging to emerge as a population-level adaptation? I explore the thesis that population dynamics is inherently chaotic, and that aging is selected for its role in smoothing demographic fluctuations. The logistic equation provides a natural vehicle for this model because it has played a central role in two sciences: Population growth in a resource-limited niche has long been modeled by the *differential* LE; and, as a *difference* equation, the LE is a canonical example of the emergence of chaos. Suppose that feedback about depleted resources generally arrives too late to avoid a wave of unsupportable population growth; then logistic population dynamics is subject to chaotic fluctuations. It is my thesis that aging is an evolutionary adaptation selected for its stabilizing effect on chaotic population dynamics.

## 1 Introduction: biological evidence demands a new theory

A broad body of biological evidence supports the notion that senescence is an adaptation, created and regulated by natural selection for its own sake[1]. However, evolutionary theorists have been skeptical of this evidence because the adaptive benefit of senescence is so diffuse, and its cost, born by the individual that carries senescence genes, so direct and immediate. For almost half a century, the biological evidence has been evaluated within the framework of prevailing theory: that senescence is a side-effect of genes selected because they enhance fertility[2,3].

But accumulating evidence makes this viewpoint more and more difficult to sustain.

- Laboratory animals bred for longevity fail to show depressed fertility[4].

- Some mechanisms of aging appear to be conserved over vast stretches of evolutionary time[5].

Demographic homeostasis & evolution of senescence

- In caloric restriction experiments, animals evince the ability simultaneously to forestall aging and increase stress resistance and immune function, even while under dietary stress[6].

- Genes have been discovered in wild populations of mice[7], worms[8] and flies[9] that appear to have no other function than to hasten the onset of senescence. When such genes are knocked out or artificially disabled, experimental animals live longer, and without apparent cost.

In light of this evidence, the need for an adaptive theory of senescence is apparent. But the group benefits generally associated with senescence (enhanced population diversity; greater efficiency of evolutionary selection) act too slowly to rescue senescence from its direct, short-term individual cost. Hence a quantitative theory for the emergence of senescence as a group-level adaptation has thus far eluded evolutionary theorists.

However, it becomes feasible to construct a model in which senescence emerges as an adaptation if account is taken of the tendency of predator-prey population dynamics to cause demographic swings, with unsustainable peaks and valleys that flirt with extinction. My thesis here is that demographic stability is a major target of selection at the population level, and that this explains the evolutionary provenance of senescence: senescence has been selected for its contribution to damping population cycles. In common with mechanisms previously explored, senescence is considered here as a form of blind, local altruism, with benefit that accrues to population members locally regardless of their kinship to the individual bearing senescence genes; however, the other mechanisms require evolutionary time scales to operate, whereas the time scale for the operation of demographic volatility is far shorter. (For example, population diversity may be increased by senescence in a few generations, but diversity itself carries no direct fitness benefit. Diversity can enhance the probability that good gene combinations emerge, and spread through a population. Only then has senescence produced a population with a fitness advantage.) Demographic instability can be lethal to a population within a few generations, and this is the primary reason that it provides a plausible path to selection of senescence genes, while population diversity and enhanced efficiency of selection do not.

**2.1 Demographic Stability in Nature**

Many ecosystems are observed to be tolerably stable, persisting over many generations with limited levels of population volatility that do not threaten constituent species with extinction. But there is no *a priori* necessity that this should be the case. Indeed, naive models of multispecies population dynamics (e.g. Lotka-Voltera equation[10]) predict deep population swings that are exceptional in our observations of nature. It is my premise that demographic stability is an evolved feature of ecosystems, deriving from selective local extinctions. Co-evolution of symbionts, territoriality, and predatory and reproductive restraint may be examples of selection related to demographic stability.

**2.2 The Logistic Equation**

The logistic equation is the oldest and simplest model of a population's approach to a steady-state level.

$$\frac{d\ln(x)}{dt} = b(1 - x/x_{ss})$$

where $d\ln(x)/dt$ is the logarithmic population growth rate, $b$ is the maximal growth rate in the absence of intraspecific competition, and $x_{ss}$ is the steady state population level. It is well-known that populations governed by the logistic equation are extremely well-behaved: $x$ approaches $x_{ss}$ asymptotically from either above or below, without overshooting[10].

But the logistic equation is equally prominent in another context entirely: as a difference equation, it is the canonical example used to study dynamic chaos. The behavior of the logistic equation with finite time increments may be either smooth or chaotic, depending on the size of $\Delta t$. For small $\Delta t$ (compared to the timescale $1/b$), the behavior is very much like the differential equation; for larger $\Delta t$, there are cycles in which $x$ overshoots $x_{ss}$, and if $\Delta t$ is increased further, the behavior undergoes a transition to dynamic chaos, such that $x$ jumps wildly about $x_{ss}$ from one time step to the next[11].

Of course, natural population systems cannot afford this kind of dynamic; they would soon fluctuate to levels too low for the population to ever recover. But from a theoretical perspective, the behavior with finite $\Delta t$ may be a more plausible model for the behavior of real populations than the smooth differential version.

## 3 Why do we expect real ecosystems to provide delayed feedback?

Steady-state population level reflects limits to the size of a population that can be supported in a given ecosystem. For animal species, one limit is set by the rate at which species lower on the food chain can regenerate and provide nourishment. Other limits come from predator species (higher on the food chain) that may bloom as their prey approach high densities, from limited physical space for habitat, and from diseases that thrive in crowded conditions. Each of these limits requires time to feed back to either birth or death rates in the target species. If the population of species B grows to the point where it depletes stores of species A on which it depends, then species A may require multiple B generations to recover. And the growth of predator (or parasitic) population C may require many generations to adjust to a higher B population level, so that it can be an effective restraining force. Once the C population has been established at a high level, it may continue to deplete population B long after B has shrunk below its steady-state level. So the time-delayed logistic equation is a plausible model for population dynamics.

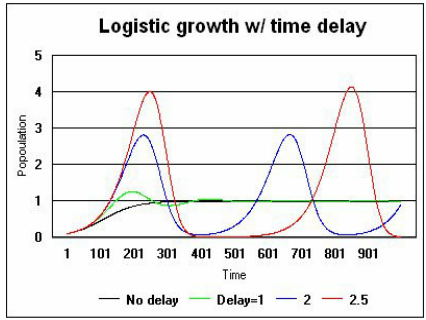

**Logistic population growth with delayed feedback**

For small time delay, population approaches a steady-state level; but for larger time delays, population is subject to cycles of ever greater severity. At delay levels that cannot be ruled out in nature, the population drops exponentially close to zero, and remain there for a substantial fraction of the cycle.

For large time delays, the shape of the population curve is reminiscent of Lotka-Voltera population dynamics, rising in a spike, then dropping exponentially close to zero for a portion of each cycle.

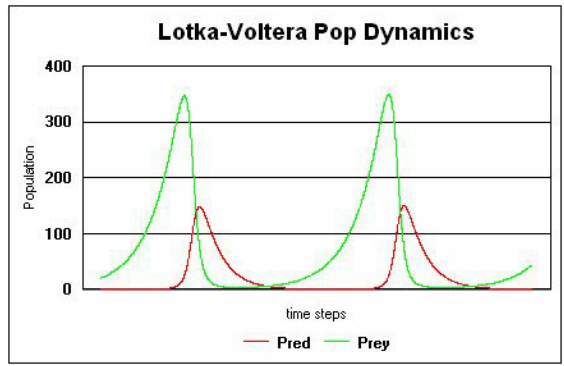

## 4 Individual-based model

Our model for a population dynamic in which senescence can evolve is an individual-based version of the time-delayed logistic equation. In each time step, each individual has a constant probability $b$ of reproducing. The act of reproduction simply creates a copy of the individual, with zero age. In each time step, each individual suffers

Demographic homeostasis & evolution of senescence a probability of death proportional to $(1-x/x_{ss})$, where population $x$ is measured at a time $\Delta t$ before the present.

This simple situation is sufficient to observe the salutary effect of senescence on population volatility; however, to actually see senescence genes selected, a population structure must be imposed. Senescence cannot evolve in a panmictic population.

Results from the un-structured model show a population volatility that increases with the time delay. If the mortality rate is increased simply by increasing the constant that multiplies $(1-x/x_{ss})$, this decreases the steady state population without adding to stability. If an extra term is added for probability of "accidental death" unrelated to crowding, this is equivalent to decreasing the birth rate $b$, and again the steady state population is reduced without any benefit to volatility. But true senescence is the increase of mortality with age. If a Gompertz term is included in the model, so that the probability of death, already proportional to collective crowding, also grows exponentially with age of the individual, then population volatility is damped, and the demography can be stabilized at somewhat higher values of the time delay than are possible without senescence.

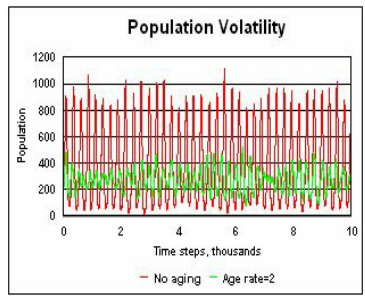 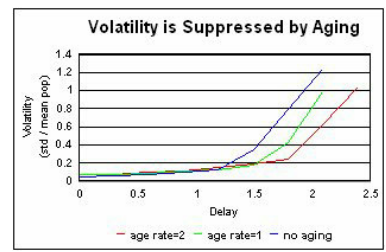

**Aging damps population fluctuations**

Population volatility (measured by the ratio of standard deviation to average population) increases with time delay. But adding Gompertz aging to the model has a stabilizing effect. (Time delays and age rates are in natural units of the population's exponential growth period for free expansion.)

## 5 Adding spatial structure: evolving senescence

To see genes for senescence selected over genes for no senescence requires a structured population. One kind of model divides the population into groups connected by low levels of migration, enabling the groups to maintain different gene frequency profiles; in another kind of model, the population is arrayed one-per-site on a viscous grid, and group structure is an emergent property. The first kind of model *must* succeed in evolving senescence, at least for a proper choice of parameter values, because there are values of the *delay* for which populations without senescence are too volatile to persist. Sites where the population is dominated by non-aging individuals will fluctuate to extinction, and migrants from neighboring sites will bring in "founders" that re-seed the site with aging types. Results from this group-array model will be presented elsewhere.

Meanwhile, I explore here an example of the second type, based on a viscous grid. On a Cartesian grid of 256*256 sites, each site may be empty, or may be occupied by an individual. Individuals have two genes: one for a constant mortality *m*, independent of age; one for a Gompertz rate of aging, *g*. In each time step, an individual has a constant probability *b* of reproducing into an empty neighboring site (if a site vacancy exists). There is also a probability of death, equal to *m* plus a *crowding* factor, all multiplied by the Gompertz aging factor.

$$mortality = (m + c * crowding)\exp(g * age)$$

The constant *c* scales the contribution of *crowding* to mortality. How should a local value of *crowding* be measured? I have used a sum over occupancy of neighboring sites, regarding each site as the center of its own neighborhood. Contributions to crowding are weighted by $1/r^2$, where $r^2=\Delta x^2+\Delta y^2$ is the Pythagorean distance between the central site and the site whose contribution is being computed:

$$crowding = \Sigma \frac{1}{\Delta x^2+\Delta y^2}$$ (with summation over occupied sites at time *delay* before the present)

I find that with *delay* set to zero, a uniform, stable level of occupancy can be maintained without aging. When *g* is left to evolve, it assumes a low value such that 3% of all deaths are senescent deaths.

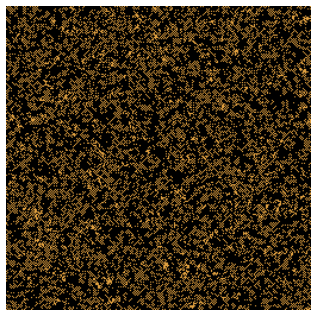

**Population modulation with instantaneous feedback**

When *delay* is set to zero, population regulates itself handily, and senescence is not necessary. Population distribution is uniform on scales larger than a few grid sites.

However, when *delay* is increased, the population wants to oscillate with a period equal to *delay*, but different regions of the grid find their own schedules. At any given time, regions of dense occupancy exist as islands in seas of empty sites. For values of *delay* that are sufficiently high, fluctuations may extinguish the population entirely. For intermediate values of *delay*, the population grows into areas where the memory of a low population density persists, only to fade from places where there is a memory of high density.

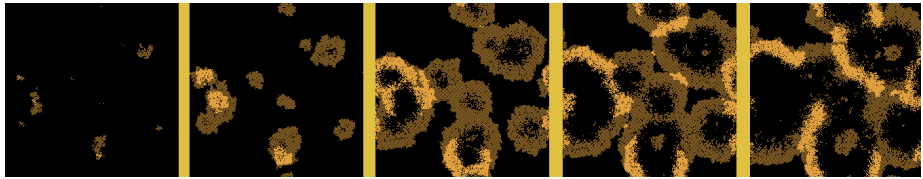

**Population moves into regions where there is a memory of low density**

After a near-extinction, population grows into space where memory of a low population density persists. Regions of high density expand into surrounding space, but then they are emptied from the inside out, as the memory of high densities at the core extinguishes population in an expanding hole.

When *g* and *m* are freed to evolve with large values of *delay*, *m* falls to zero, but *g* evolves to high levels, such that a substantial fraction of all deaths are senescent deaths. (This is my criterion for the evolution of a significant level of aging.)

**Evolved rate of aging rises with time delay**

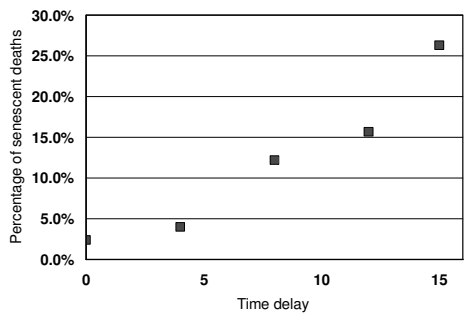

**Aging evolves in a viscous population governed by delayed logistic population growth**

Aging evolves as a mechanism to stabilize population dynamics, as regions without aging experience wide extremes, and fluctuate to extinction. Aging is programmed as an exponential (Gompertz) increase of death rate with age, beginning at an age $2/b$, where $b$ is the rate of reproduction. The exponential rate of increase is governed by a gene *g*. For large time delays, I observe evolved rates of aging sufficient to account for a substantial proportion of all deaths.

For small values of delay (≲4), the evolutionary process seems stochastic and indeterminate. A range of values for *agerate* seems equally able to compete with no aging.

## 6 Summary

Experimental evidence compels the conclusion that aging is an adaptation; but theorists have been reluctant to accept this paradigm because aging carries a cost to the individual, and no group-level adaptive value for aging has been identified that acts promptly and powerfully enough to counterbalance the individual cost. I suggest here that aging has been selected for the contribution it makes to demographic stability.

The time-delayed logistic equation is an appropriate model for population fluctuations, rooted as it is both in population genetics and chaos theory. In order to turn the logistic equation into an individual-based model that can support the evolution of senescence, two things are necessary: a spatially structured population, and a reasonable definition of crowding. Providing these in a straightforward way, I find that a gene for aging emerges readily, without need for special assumptions or tuning of parameter values. The principal assumption is that the time delay for population feedback is sufficiently high that demographic instabilities pose a significant risk of extinction. Under these circumstances, senescence (defined as an increase in mortality with age) stabilizes population swings in a way that neither a lower birth rate nor a higher mortality rate can. (The latter led only to lower steady-state population levels, but not decreased volatility.)